\documentclass[aps,prl,epsfig,floats,twocolumn,amssymb,amsmath,showpacs,footinbib]{revtex4-1}

\usepackage{color}

\usepackage{graphicx}

\begin{document}

\title{Length scale dependence of DNA mechanical properties}

\author{Agnes Noy}
\affiliation{Rudolf Peierls Centre for Theoretical Physics, University of Oxford, 1 Keble Road, Oxford, OX1 3NP, UK}
\author{Ramin Golestanian}
\affiliation{Rudolf Peierls Centre for Theoretical Physics, University of Oxford, 1 Keble Road, Oxford, OX1 3NP, UK}

\date{\today}

\begin{abstract}
Although mechanical properties of DNA are well characterized at the kilo base-pair range, a number of recent experiments have suggested that DNA is more flexible at shorter length scales, which correspond to the regime that is crucial for cellular processes such as DNA packaging and gene regulation. Here, we perform a systematic study of the effective elastic properties of DNA at different length scales by probing the conformation and fluctuations of DNA from single base-pair level up to four helical turns, using trajectories from atomistic simulation. We find evidence that supports cooperative softening of the stretch modulus and identify the essential modes that give rise to this effect. The bend correlation exhibits modulations that reflect the helical periodicity, while it yields a reasonable value for the effective persistence length, and the twist modulus undergoes a smooth crossover---from a relatively smaller value at the single base-pair level to the bulk value---over half a DNA-turn.
\end{abstract}


\maketitle
Single molecule experiments have characterized DNA in the kilo-base-pair (kbp) range as an ideal elastic rod with a persistence length of 50 nm, twist elastic constant between 80 and 120 nm, and a stretch modulus of 1100-1500 pN \cite{busta92,busta96,croquette96,nelson97,gross2011twist}. However, recent advances in experimental techniques that are capable of probing short fragments have provided new data on DNA mechanical properties that are sometimes dissimilar to the single-molecule results \cite{05Dietler,nelson06,fretsaxs,mastroianni2009probing}, suggesting that DNA elastic properties might depend on fragment length. Atomic Force Microscopy (AFM) measures have revealed an unexpectedly high number of large bends compared as compared with the predictions of the wormlike chain (WLC) model \cite{nelson06}, while fluorescence resonance energy transfer (FRET) technique combined with small-angle x-ray scattering (SAXS) provided a persistence length that was a factor of two smaller than the kbp range value \cite{fretsaxs}. More controversial was the recent SAXS measurement of stretch deformability using the end-to-end distance of fragments shorter than 40 bp, which provided a value one order of magnitude lower than previous estimations, presumably due to a cooperative effect  \cite{mathew2008remeasuring,becker2009comment,mathew2009response}. Despite the significance of DNA elasticity at this scale, a comprehensive theoretical picture that relates these findings with the detailed atomistic structure of DNA is still lacking.

\begin{figure}
\centerline{\includegraphics[width=1.00 \columnwidth]{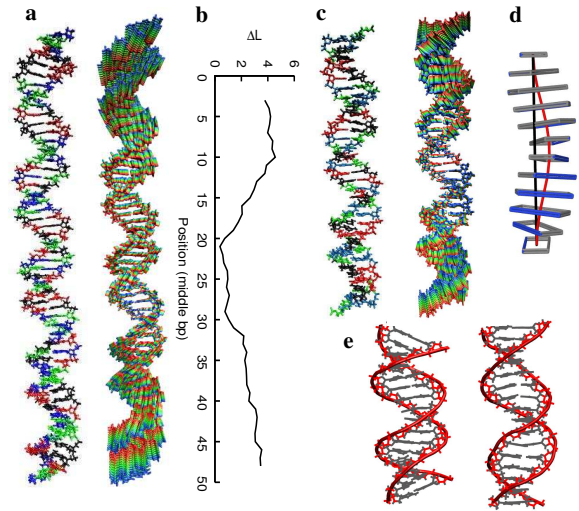}}
\caption{56-mer (a) and 36-mer averages (c) together with the corresponding end-stretching essential modes (see Movie S1). (b) Molecular position dependence of the local strain for the end-to-end distance ($\Delta L$) occurred at the 56-mer end-stretching mode using sub-fragments of 3 bp length. (d) Side view of a regular straight DNA turn made using dimer average values from our simulations (shift=0.0, slide=-0.5 and rise=3.3 \AA; tilt=0, roll=3, twist=33 degrees) where end-to-end (black) and contour length (red) are differentiated. (e) Snapshots from the ``zigzag'' essential mode for the 56-mer central 16 bp (see Movie S2).}\label{figure1}
\end{figure}

In this paper, we study how the elastic properties of DNA evolve with the length scale, from the scale of dinucleotides to near four helical turns. We start from an atomistic description of DNA elasticity, and perform a systematic program of probing the geometric correlation at different length scales using the trajectories from molecular dynamics (MD) simulation. We aim to examine the potential effect of the asymmetric double-helical structure on the elastic response of the molecule and find out at which scale DNA starts to behave as an isotropic elastic rod with the elastic constants approaching the large-scale experimental values. To this end, we have simulated two randomly sequenced DNA molecules (see Fig. 1). Using all possible internal fragments with a particular length and corresponding to different sequences, we are able to trace the generic mechanical properties of double-stranded DNA at that length scale. The statistics is further helped by using the two different fragments with completely unrelated random sequences. We examine how elastic probes at small scales could be sensitive to the definition used in measuring them, and use this to shed light on how the results of experiments at small scales should be analyzed and compared in view of this definition-dependence.

We built a 56-mer (CGCGATTGCC TAACGGACAG GCATAGACGT CTATGCCTGT CCGTTAGGCA ATCGCG) and a 36-mer (CGCGATTGCC TAACGAGTAC TCGTTAGGCA ATCGCG), containing together at least 5 copies for each of the 10 unique dinucleotide steps, by using the AMBER8 package with parm99+parmbsc0 force field \cite{amber8,bsc0}. Then, structures were solvated and neutralized using GROMACS-4 \cite{gromacs} with dodecahedron boxes ($\sim$226000 and $\sim$71000 SPC/E water molecules \cite{spce}, respectively) and with 110 and 70 Na$^{+}$ counterions placed randomly, respectively \cite{dang94}. Systems were energy-minimized, thermalized ($T=298$ K) and equilibrated using a standard protocol \cite{perez08,noy2010chirality} followed by a final re-equilibration of 20 ns \cite{mazur06}. The final structures were subject to 100 ns of productive MD simulation at constant temperature (298 K) and pressure (1 atm) using periodic boundary conditions and particle mesh Ewald \cite{ewald}.

The DNA geometric variables for different length scales are obtained by extending the algorithm of the widely used 3DNA program \cite{3dna} to describe the geometry between 2 base-pairs spaced by an increasing number of nucleotides. Thus, the identical configuration for an arbitrary base-pair $i$ is specified by giving the location of a reference point $r_i$ and the orientation of a right-handed orthonormal frame where $\hat{x}_i$ points to the major groove and $\hat{z}_i$ is the unit vector tangent of the local direction. Then, the deformation of the double helix is characterized by the bending angle $\theta=\cos^{-1}(\hat{z}_i\cdot\hat{z}_{i+n})$, the end-to-end distance $L=|r_{i+n}-r_i|$, and the contour length $L_0=\sum_{i}^{n-1} |r_{i+1}-r_i|$ (Fig 1c). Twist as well as the 2 bending components, tilt and roll, are calculated using the concept of the mid-step triad as defined in 3DNA (Fig. S1 \cite{Note1}).

Elastic constants at different fragments lengths are evaluated as the diagonal terms of the elastic matrix $F$ calculated by \cite{olson98}:
\begin{equation}\label{invcov}
 F=k_{\rm B}TNb\;V^{-1},
\end{equation}
where $V$ stands for the 4x4 covariance matrix of DNA deformation variables (roll, tilt, twist, and stretch which can be either the end-to-end distance or the contour length) for that particular length. Fragment length is defined by $N$ dinucleotide steps with rise $b=0.34$ nm. In addition, the diagonal terms of $V^{-1}$ can be understood as the reciprocal of the partial variances, which measure the residual variance associated with given deformations after removing the linear effects caused by the other variables \cite{Whittaker,Note1}.

\begin{figure}
\includegraphics[width=1.00 \columnwidth]{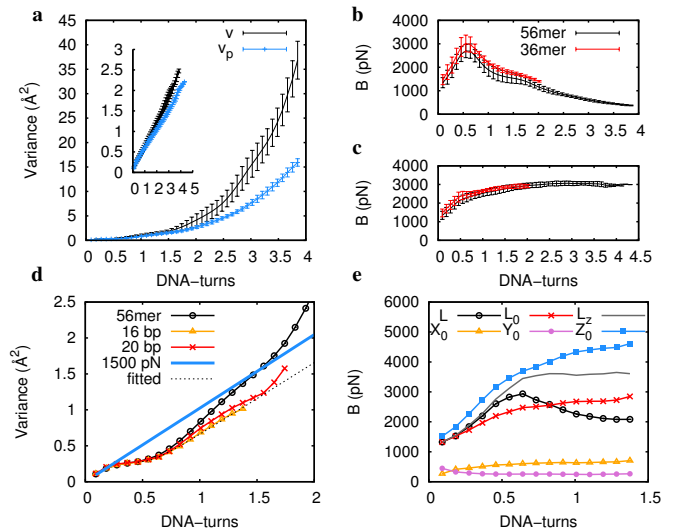} \\
\caption{(a) End-to-end $L$ as well as contour-length $L_0$ (inset) raw ($v$) and partial variances ($v_p$) as extracted from the 56-mer trajectory. Stretch modulus $B$ associated with the $L$ (b) and to the $L_0$ (c) from MD trajectories. Values reported here are averages over all possible sub-fragments with a particular length, and the corresponding standard deviations are given as error bars. (d) End-to-end partial variances extracted from the 56-mer trajectory considering the whole molecule, the central 16 and 20 bp, predictions for a 1500 pN stretch modulus and the fitted curve. (e) $B$ calculated for the 56-mer's central 16 bp associated with $L$, $L_0$, the corresponding added rise ($Z_0$), added lateral shear (slide $Y_0$ and shift $X_0$), and with $L$ of the trajectory with the ``zigzag'' essential mode filtered out ($L_z$). Note that $B$ associated to added rise and lateral shear is obtained by full variances instead of partial variances.}
\end{figure}

We probed the stretching response of DNA using two different definitions, corresponding to the contour-length and the end-to-end distance, for a better understanding of how various experimental results can be interpreted. The end-to-end distance is known to have a contribution from bending  \cite{becker2009comment}, which we eliminate using Eq. (\ref{invcov}). We find that the end-to-end partial variance still presents a faster-than-linear growth as a function of the fragment length (Fig. 2a). Since for completely independent random fluctuations we expect the cumulative variance to simply add up for neighboring segments and grow linearly, the super-linear growth implies a certain correlation between the neighboring steps and as a consequence a cooperative behavior \cite{mathew2008remeasuring}. The corresponding elastic modulus exhibits a non-monotonic behavior upon increasing the length scale (Fig. 2b): it increases as the length scale is increased from single base-pair level to approximately half DNA-turn, above which a softening of nearly one order of magnitude is observed. This is similar to the trend observed in recent SAXS studies \cite{mathew2008remeasuring}. On the other hand, when using the contour-length definition, the variance shows a non-cooperative (nearly) linear dependence on length (inset of Fig. 2a), with a corresponding stretch modulus that starts from a value of 1200 pN and approaches a plateau of around 3000 pN (Fig. 2c).

To find out the origin of the stretch modulus softening we performed essential dynamics analysis \cite{amadei93} over both trajectories using the GROMACS package. We found an essential mode that primarily corresponds to stretching the end sections (Fig. 1 and Movie S1), which is capable of transforming the nearly linear contour-length variance into the observed cooperative form when the trajectories are projected onto it (\emph{i.e.} when the contour length variance is filtered from uncorrelated vibrations; see Fig. S2). We further focus on the central 16 bp of the 56-mer, to probe other contributions than the end effects, and using the duplex model with partial correlations proposed by Matthew-Fenn \emph{et al.} \cite{mathew2008remeasuring} we find evidence of cooperativity with a crossover to linear behavior around 8 bp and a final slope of 1518 pN (variance per base step of 0.027 \AA$^2$ $\pm$ 0.002 and reduced $\chi^2=0.16$) (Fig. 2d). Thus, our results indicate that the experimentally soft stretch modulus measured by the end-to-end distance could be caused mainly by end effects, and that the stretch modulus would reach a plateau around 1500 pN in the internal fragments of a much longer DNA molecule.

A remaining puzzle is why end-to-end and contour-length stretch definitions provide such different elastic responses after the bending contribution is eliminated. To shed light on this, and the high stiffness peak around half DNA-turn, we calculate the stretch modulus associated to the 3 components of the contour-length, namely, the longitudinal component, rise, and the two lateral shear components, slide and shift \cite{Note1} of the central 16 bp of the 56-mer. We observe that rise is much stiffer than lateral shear, due to the strong base-stacking interactions \cite{noy2010chirality}, and that the contour-length effective elastic modulus can be thought of as an average value of the three local force constants (Fig 2e). On the contrary, the end-to-end distance is predominantly influenced by rise at short length scales, which is why it exhibits a relative stiffening at scales shorter than half a DNA-turn (Fig. 1c).

To understand the cooperative behavior of the end-to-end distance at 5-10 bp scale, we performed essential dynamics analysis over the central fragment. We found that among the first ten essential components, there was only one mode that was capable of differentiating between the two definitions (Fig. 1d and Movie S2); the cooperativity would be removed only when this mode was filtered out from the trajectory (Fig. 2e). The essential mode can be characterized by the change in the inclination and displacement of the bases relative to the global molecular axis (Fig. S3 \cite{Note1}). Therefore, our results suggest that whereas the contour length would be influenced by uncorrelated local fluctuations that result in a monotonic increase in the variance (and a plateau in the stretch modulus), the end-to-end fluctuations would capture a larger number of structured modes where several base-pairs would move in a coordinated way.

\begin{figure}
\includegraphics[width=8.6cm]{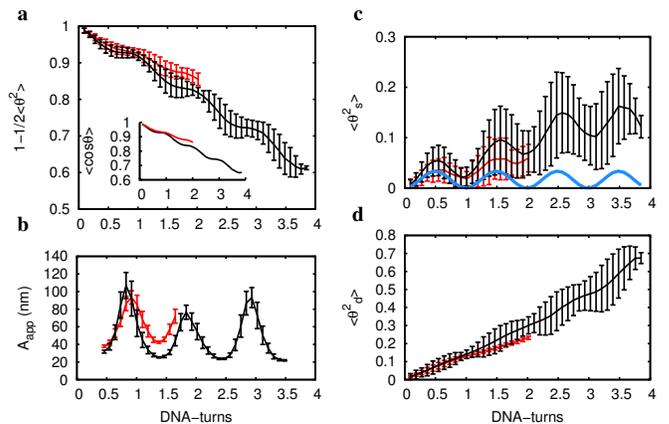}
\caption{(a) Directional decay of the 56-mer (black) and the 36-mer (red) MD trajectories. The inset shows the exact cosine function between base-pairs axial vectors, while the rest of the plots use the quadratic approximation (see Eq. \ref{bend}). (b) Apparent persistence length $A_{\rm app}$ calculated by a linear fit of the local directional decay (2 parameters, 5 degrees of freedom, all reduced $\chi^2$'s were under 0.2) where the asymptotic standard errors of the fitted persistence length are given as error bars. The static (c) and the dynamic (d) contributions to the fluctuations of the bending angle $\theta$ (see Eq. \ref{ThetaFlucTot}) calculated from the MD trajectories, as well as that of the regular straight B-DNA (in thick blue) for comparison.}
\end{figure}

According to the WLC model, the persistence length $A$ is quantified through the orientational correlation between two tangent vectors, $\hat{z}_i$ and $\hat{z}_j$, that are separated by a distance $L_{i,j}$ along the DNA as $\left< \cos \theta_{i,j}\right>={e}^{-L_{i,j}/A}$, where we have defined $\theta_{i,j}$ as the  bending angle. For a sufficiently weakly bending rod, a quadratic approximation can be used in the power series to yield:
\begin{equation}\label{bend}
\left< \cos \theta_{i,j}\right>\cong 1-\frac{1}{2}\left<\theta_{i,j}^2\right> \equiv 1-\frac{L_{i,j}}{A},
\end{equation}
where $\left<\theta_{i,j}^2\right>$ can be obtained by averaging using the MD trajectories and the persistence length $A$ can be read off using the above definition. The result of this calculation is shown in Fig. 3a (in which the inset confirms the validity of the above approximation). We observe that the correlation between local tangent vectors does not decay uniformly with distance; it exhibits a DNA-turn periodicity \cite{mazur06,olson88}.
In particular, when a fragment length is between $n$ and $n+\frac{1}{2}$ helical turns, it would appear that the orientation memory is lost more rapidly, whereas for lengths between $n+\frac{1}{2}$ and $n+1$ turns the DNA fragment would appear to behave as a stiffer polymer. The nonlinear behavior shows that the value of the persistence length will depend on the definition used to deduce it from the decay of the bending correlations. Using a {\em punctual-decay} \cite{nelson06} as the definition, for example, we obtain a value of around 34 nm for the persistence length of a half DNA-turn fragment, whereas if we consider a full DNA-turn fragment we obtain a value of around 49 nm. The difference between the minimum and the maximum average values decreases as we increase the length, and an overall average slope yields a persistence length of 43 nm (see Fig. S4). If a local definition is used to calculate the slope (as in \cite{fretsaxs} where fragment lengths from 17 to 21 bp are used), and thereby the apparent persistence length $A_{app}$, these variations will be much more pronounced, and range from 20 nm to 100 nm (Fig. 3b and Fig. S4).

The behavior of the bending correlations can be understood in terms of the static (Fig. 3c) and dynamic (Fig. 3d) contributions to the bending angle fluctuations:
\begin{equation}\label{ThetaFlucTot}
\left<\theta^2\right>=\left<\theta^2_s\right>+\left<\theta^2_d\right>.
\end{equation}
The periodicity observed in the static bending correlations  (Fig. 3c) comes from the structure of DNA at the dinucleotide level, as observed from the analysis of experimental structures \cite{dickerson97, Note2}. Averaging over all possible dinucleotide steps of the studied fragments, we observe a net bending of 3 degrees towards the major groove, in agreement with previous all-atom simulations, and experimental x-ray and NMR structures (roll $\approx$ 3, 1.5 and 2.5 degrees respectively) \cite{olson98,perez08}. When the directional correlations between base-pairs are evaluated for the corresponding regular straight B-DNA (Fig. 1c) a similar periodicity is found (Fig. 3c). The net linear increase in $\left<\theta^2_s\right>$ with length comes from a (quenched) random distribution of sequence-dependent static bends obtained through the MD average structure, from which a static persistence length can be inferred as $A_s=216$ nm (Fig. 3c). Then, the dynamic persistence length can be extracted from Eq. \ref{ThetaFlucTot} which yields $A_d=54$ nm (Fig. 3d, \cite{Note1}). The static and the dynamic persistence lengths
are combined using $1/A=1/A_s+1/A_d$ \cite{schellman} to obtain an average persistence length of $A=43$ nm, which is compatible with a direct average fit to the full bending angle correlation decay (Fig. 3a).

Moreover, we find that this value contains a contribution from twist, which we can remove by using the partial variance instead of the full variance \emph{i.e.} by using $A_d^{-1}=1/2(A_{\rm roll}^{-1}+A_{\rm tilt}^{-1})$ \cite{Note1,norouzi2008effect,landau}. The partial-variance method yields an increased dynamic persistence length of $A_d=67$ nm, and a resulting average persistence length of $A=51$ nm, which is in agreement with single-molecule experiments \cite{busta92}. Thus, our results suggest that the higher bending flexibility of DNA detected by AFM imaging \cite{nelson06} and FRET/SAXS \cite{fretsaxs} as compared to the force-extension measure could be influenced by the length-dependent static curvature as well as the use of a local definition for the persistence length. However, we would like to point out that some of the observed effects in Ref. \cite{nelson06} could also be produced by the presence of Mg$^{\mathrm{2+}}$, which is known to decrease the persistence length \cite{baumann97}.

\begin{figure}
\includegraphics[width=0.7 \columnwidth]{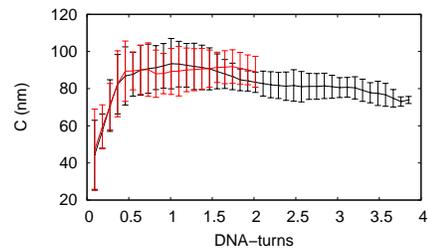}
\caption{Torsion elastic constant $C$ calculated by the inverse covariance matrix (Eq. \ref{invcov}) using the 56-mer (black) and the 36-mer (red) MD trajectories.}
\end{figure}

Finally, the calculated torsion elastic modulus shows a crossover from the relatively soft value of around 50 nm at the single base-pair level to a large-scale asymptotic value of around 80 nm, which is consistent with the single-molecule experiments \cite{croquette96,nelson97}. The crossover is monotonic and occurs below 1 DNA-turn (Fig. 4), which suggests that as regards to the twist response, DNA would behave like a elastic rod with a single elastic constant at all scales except for fragments shorter than a full DNA turn that are roughly speaking softer by a factor of two and can undergo relatively large twist conformational changes.

In summary, we find that the elastic properties of DNA depend on the length scale with various degrees of sensitivity for the different modes of deformation. Our study suggests that a simple description of the conformational response of DNA that uses a few elastic moduli can be used for short fragments of DNA provided their length-scale dependencies are taken into consideration and care is taken on how these quantities are defined. The crossover from single base-pair level to bulk elastic behavior occurs typically within one helical turn of DNA, except for the relatively more long-ranged end-effects observed for the stretching response of the molecule. The observed length-scale dependence could have important implications in such fundamental processes as DNA-protein interactions, and DNA looping or packing inside the cell. Our study could be extended to probing the length scale dependence of the various aspects of the elastic response of DNA, such as the relative contribution of electrostatic and non-electrostatic interactions to the various elastic moduli \cite{11JACSPapoian}.

We would like to thank M. Orozco and W.K. Olson for useful discussions and comments, S.A. Harris for reading the manuscript, and the Red Espanola de Supercomputacion for computational resources. A.N. is an EMBO long-term fellow.

\end{document}